\documentclass[aps,prl,reprint,nofootinbib,superscriptaddress]{revtex4-2}

\usepackage[T1]{fontenc}
\usepackage[utf8]{inputenc}
\usepackage{amsmath,amssymb,amsfonts}
\usepackage{graphicx}
\usepackage{bm}
\usepackage{hyperref}
\usepackage{tikz}
\usepackage{xcolor}
\usepackage{orcidlink}
\usetikzlibrary{arrows.meta,calc}

\hypersetup{
  colorlinks=true,
  linkcolor=blue,
  citecolor=blue,
  urlcolor=blue
}

\begin{document}

\title{Massless Islands in Wedge Holography}

\author{Naman Kumar\,\orcidlink{0000-0001-8593-1282}}
\affiliation{Department of Physics, Indian Institute of Technology Gandhinagar, Palaj, Gujarat, 382355, India}
\email{namankumar5954@gmail.com,naman.kumar@iitgn.ac.in}

\date{\today}

\begin{abstract}
Entanglement islands are usually easiest to realize in doubly holographic models with massive gravitons or non-gravitating baths.  In wedge holography, by contrast, Neumann boundary conditions on both branes give a normalizable massless graviton, but the island saddle of the purely geometric Ryu-Takayanagi (RT) problem collapses to the horizon.  Negative Dvali-Gabadadze-Porrati (DGP) terms can restore nontrivial islands by changing the endpoint condition of the extremal surface, but this branch contains a massive ghost. We propose a different semiclassical mechanism.  The minimal wedge obstruction is a statement about the purely geometric RT functional, not a no-go for quantum extremal surfaces once the healthy Neumann wedge sector is coupled to additional quantum matter at the corner.  We keep the wedge gravitational action free of DGP terms and add a unitary defect conformal field theory localized at the codimension-two corner.  This sector is distinct from the standard corner
CFT dual to the undeformed wedge, whose entropy is already represented by
the wedge RT area; hence no double counting is involved. The entropy of this additional defect sector is the ordinary matter-entropy term required by the QES prescription, rather than an ad hoc correction to the RT area.  If the defect theory is holographic, this matter entropy can be evaluated by an auxiliary bulk Ryu-Takayanagi surface.  We show that when the wedge endpoint position determines the defect entangling region, the auxiliary area can have the opposite variation to the wedge area while all couplings and central charges remain positive.  A local endpoint model exhibits an isolated stable saddle and its late-time dominance over the Hartman-Maldacena (HM) surface.  The quantum extremality condition then replaces the pure orthogonality condition and allows a non-horizon island saddle in a long-range, massless, ghost-free gravitational theory.  This demonstrates that the obstruction to massless islands in minimal wedge holography is not masslessness itself, but the absence, in the minimal geometric RT problem, of a healthy matter-entropy term capable of balancing the horizon-minimizing area variation.
\end{abstract}

\maketitle

\emph{Introduction.---} The island formula has provided a concrete semiclassical mechanism for recovering Page curves in black hole evaporation and eternal black hole setups \cite{Penington:2019npb,Almheiri:2019psf,Almheiri:2019hni,Almheiri:2020cfm}.  In higher dimensions, most controlled models use double holography, where the gravitational system is coupled to a non-gravitating bath or to branes supporting massive gravitons \cite{Karch:2000ct,Takayanagi:2011zk,Almheiri:2019psy,Geng:2020qvw}.  This has led to a puzzle: can islands exist in a theory with genuinely long-range, massless gravity?

\begin{figure}[t]
\centering
\resizebox{\columnwidth}{!}{%
\begin{tikzpicture}[
    x=1cm,y=1cm,
    >=Latex,
    font=\scriptsize,
    panel/.style={draw, rounded corners=2pt, line width=0.4pt},
    arrow/.style={->, line width=0.5pt},
    wedge/.style={fill=blue!10, draw=orange!80!black, line width=0.55pt},
    brane/.style={fill=orange!10, draw=orange!80!black, line width=0.55pt},
    corner/.style={draw=green!50!black, line width=0.6pt},
    ent/.style={draw=green!50!black, dashed, line width=0.5pt}
]

\def\pw{2.15}
\def\ph{2.25}

\begin{scope}[shift={(0,0)}]
\draw[panel] (0,0) rectangle (\pw,\ph);
\node[anchor=west] at (0.03,2.48) {(a) $(d{+}1)$ bulk};

\coordinate (L1) at (0.28,1.86);
\coordinate (L2) at (0.28,0.42);
\coordinate (R)  at (1.78,1.14);

\filldraw[wedge] (L1)--(R)--(L2)--cycle;
\draw[dashed, gray, line width=0.35pt] (0.18,1.86)--(0.18,0.42);

\node[text=blue!65!black, font=\small] at (0.88,1.14) {$W$};
\node[text=orange!80!black] at (0.95,1.65) {$Q_1$};
\node[text=orange!80!black] at (0.95,0.62) {$Q_2$};
\fill (R) circle (1pt);
\node[anchor=west] at ($(R)+(0.05,0)$) {$\Sigma$};

\node at (1.07,0.15) {wedge};
\end{scope}

% Arrow
\draw[arrow] (2.25,1.14)--(2.62,1.14);
\node[font=\tiny] at (2.44,1.36) {AdS/CFT};

\begin{scope}[shift={(2.72,0)}]
\draw[panel] (0,0) rectangle (\pw,\ph);
\node[anchor=west] at (0.03,2.48) {(b) $d$ branes};

\coordinate (U1) at (0.35,1.78);
\coordinate (U2) at (0.35,1.32);
\coordinate (M)  at (0.88,1.12);
\coordinate (R)  at (1.78,1.12);
\coordinate (U3) at (1.62,1.34);

\coordinate (L1) at (0.35,0.46);
\coordinate (L2) at (0.35,0.92);
\coordinate (L3) at (1.62,0.91);

\filldraw[brane] (U1)--(U3)--(R)--(M)--(U2)--cycle;
\filldraw[brane] (L1)--(L3)--(R)--(M)--(L2)--cycle;
\draw[orange!80!black, line width=0.55pt] (M)--(R);

\foreach \y in {1.52,1.38,0.74,0.60}{
  \draw[line width=0.35pt] (0.55,\y)
  .. controls (0.64,\y+0.05) and (0.73,\y-0.05) .. (0.82,\y);
}

\node[text=orange!80!black] at (1.00,1.64) {$Q_1$};
\node[text=orange!80!black] at (1.00,0.62) {$Q_2$};
\fill (R) circle (1pt);
\node[anchor=west] at ($(R)+(0.05,0)$) {$\Sigma$};

\node at (1.07,0.15) {brane gravity};
\end{scope}

\draw[arrow] (4.97,1.14)--(5.34,1.14);
\node[font=\tiny] at (5.16,1.36) {AdS/CFT};

\begin{scope}[shift={(5.44,0)}]
\draw[panel] (0,0) rectangle (\pw,\ph);
\node[anchor=west] at (0.03,2.48) {(c) $(d{-}1)$ corner};

\coordinate (T) at (0.72,1.70);
\coordinate (B) at (0.72,0.72);

\draw[corner] (T)--(B);
\fill[green!50!black] (T) circle (1pt);
\fill[green!50!black] (B) circle (1pt);
\node[anchor=south] at ($(T)+(0,0.06)$) {$\Sigma$};

\node[text=green!50!black] at (1.45,1.36) {$\mathrm{CFT}_{d-1}$};

\draw[ent] (0.72,0.72)
.. controls (1.28,0.80) and (1.28,1.08) .. (0.72,1.16);
\node[text=green!50!black] at (1.32,0.94) {$A$};

\node at (1.07,0.15) {corner CFT};
\end{scope}

\node[align=center, font=\tiny] at (3.80,-0.38)
{$W$ geometry $\Longleftrightarrow$ gravity on $Q_1\cup Q_2$
$\Longleftrightarrow$ entanglement in $\mathrm{CFT}_{d-1}$};

\end{tikzpicture}%
}
\caption{The three equivalent descriptions in wedge holography: a
$(d{+}1)$-dimensional wedge bulk, a $d$-dimensional brane-gravity
description, and a $(d{-}1)$-dimensional corner CFT.}
\label{fig:wedge-three-descriptions}
\end{figure}
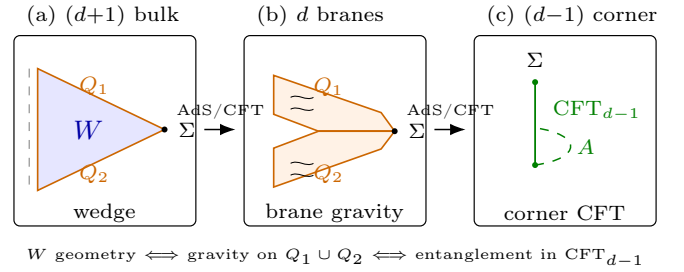

Wedge holography is an especially sharp arena for this question
\cite{Akal:2020wfl,Hu:2022ymx}. The wedge has two end-of-the-world branes
$Q_1\cup Q_2$ meeting at a codimension-two corner $\Sigma$. It is a
\emph{doubly holographic} realization in which three descriptions are related
by successive applications of the AdS/CFT correspondence
\cite{Maldacena:1997re,Witten:1998qj}: a $(d+1)$-dimensional gravitational
description in the wedge bulk $W$, a $d$-dimensional gravitational description
on the branes $Q_1\cup Q_2$, and a $(d-1)$-dimensional nongravitational CFT
description on the corner $\Sigma$. In this sense, wedge RT surfaces, brane
gravity, and corner-CFT entanglement provide equivalent descriptions of the
same system (see Fig.~\ref{fig:wedge-three-descriptions}). With Neumann boundary conditions on both branes, the brane graviton includes a normalizable massless mode.  However, in the minimal wedge model the island-phase extremal surface is forced to coincide with the horizon.  Equivalently, the free endpoint variation enforces orthogonality at the branes, and this condition rules out non-horizon island surfaces \cite{Geng:2020fxl,Geng:2021iyq}. DGP terms on the branes modify this endpoint condition and can produce massless islands \cite{Dvali:2000hr,Miao:2022mdx,Miao:2023unv}.  The difficulty is that the required DGP coefficient is negative, and the same branch contains a ghost in the Kaluza-Klein spectrum.  Thus the known geometric route to massless islands appears to pass through an unstable sector.

In this Letter we describe a different mechanism.  We do not add a DGP term.  Instead we add an explicit, unitary defect conformal field theory localized on $\Sigma$, distinct from the corner CFT that is dual to the undeformed wedge system.  Its matter entropy is included in the generalized entropy.  If the added defect CFT is holographic, this entropy is represented by an auxiliary Ryu-Takayanagi (RT) area \cite{Ryu:2006bv} in an auxiliary bulk $W_{\rm aux}$.  The key observation is simple: the auxiliary RT area need not vary with the same sign as the wedge RT area.  If the moving wedge endpoint increases the defect entangling region, the auxiliary entropy increases as the wedge surface is pushed toward the horizon.  This supplies the same kind of endpoint force that negative DGP supplied, but with positive central charge and positive auxiliary Newton constant.  This should be viewed as a maximin selection mechanism: the additional defect entropy lifts the horizon-collapse or trivial-partition problem in the same way that radion or dilaton fluctuations lift the classical RT degeneracy in $AdS_3$ wedge holography \cite{Geng:2022slq}.

\emph{Minimal wedge model and the no-island saddle.---} We work in a $(d+1)$-dimensional wedge $W$ bounded by two branes $Q_1$ and $Q_2$.  A convenient black-string geometry is
\begin{align}
 ds^2&=dr^2+\cosh^2 r\,
 \frac{dz^2/f(z)-f(z)dt^2+d\vec y_{d-2}^{\,2}}{z^2},
\nonumber\\
 f(z)&=1-z^{d-1},
\label{eq:blackstring}
\end{align}
with $-\rho_1\leq r\leq \rho_2$.  The branes are at $r=-\rho_1$ and $r=\rho_2$, while the horizon is at $z=1$.  The minimal wedge action without DGP terms is
\begin{align}
 I_{\rm wedge}={}&
 \frac{1}{16\pi G_W}\int_W d^{d+1}x\,\sqrt{-g}\,
 \bigl(R+d(d-1)\bigr)
\nonumber\\
&+\frac{1}{8\pi G_W}\sum_{a=1}^{2}
 \int_{Q_a} d^d x\,\sqrt{-h_a}\,(K_a-T_a).
\label{eq:minimalaction}
\end{align}
Here \(R\) is the Ricci scalar of the wedge bulk \(W\), \(Q_a\) with
\(a=1,2\) are the two end-of-the-world branes, \(h_a\) is the induced
metric on \(Q_a\), \(K_a\) is the trace of its extrinsic curvature, and
\(T_a\) is its tension.  We have set the AdS radius to unity.  We impose
Neumann boundary conditions on both branes.  This is the branch in which
wedge holography supports a normalizable massless graviton.

For an island-phase RT surface $\Gamma_W$ parametrized by $t={\rm const.}$ and $z=z(r)$, the geometric entropy functional is proportional to
\begin{equation}
 A_W[z]=V\int_{-\rho_1}^{\rho_2}dr\,
 \frac{\cosh^{d-2}r}{z(r)^{d-2}}
 \sqrt{1+\frac{\cosh^2r\,z'(r)^2}{z(r)^2 f(z(r))}},
\label{eq:wedgearea}
\end{equation}
where $V=\int d^{d-2}y$.  Since $0<z\leq 1$ outside the horizon and $f(z)\geq0$, one obtains
\begin{equation}
 A_W[z]\geq V\int_{-\rho_1}^{\rho_2}dr\,\cosh^{d-2}r\equiv A_{\rm hor} .
\label{eq:ineq}
\end{equation}
Thus the purely geometric island surface is the horizon.  The same conclusion follows from the endpoint variation: for a freely moving endpoint $z_a=z((-)^a\rho_a)$, the boundary term in $\delta A_W$ vanishes only when the surface ends orthogonally on the brane,
\begin{equation}
 z'_a=0 .
\label{eq:orthogonal}
\end{equation}
In the black-string wedge this condition is strong enough to leave only the horizon saddle.  This is the minimal wedge no-island result.

The minimal wedge obstruction is a statement about the purely geometric RT
functional \(A_W/4G_W\).  It does not by itself exclude nontrivial QES saddles
in a semiclassical wedge theory coupled to additional quantum matter.  Once a
localized corner sector is included, the appropriate functional is
\begin{equation}
        S_{\rm gen}
        =
        \frac{A_W}{4G_W}
        +
        S^{\rm extra}_{\Sigma}.
\label{eq:qes-semiclassical-corner}
\end{equation}
The entropy \(S^{\rm extra}_{\Sigma}\) is not an ad hoc correction to the RT
area; it is the ordinary matter-entropy term required by the QES prescription.
The role of the added defect sector is therefore to supply a healthy positive
entropy variation that competes with the horizon-minimizing wedge area, while
leaving the Neumann wedge gravity sector on the massless ghost-free branch.

\emph{Defect deformation.---}
We now deform the minimal wedge theory by adding an explicit unitary defect
CFT localized on the corner,
\begin{equation}
 I=I_{\rm wedge}+I_{\Sigma},\qquad
 I_{\Sigma}=I_{\rm CFT}^{\rm extra}[\psi,\sigma]+{\color{blue}I_{\rm int}} .
\label{eq:defectaction}
\end{equation}
Here \(\sigma\) is the induced metric on \(\Sigma\), and \(\psi\) denotes
the additional defect degrees of freedom. The term
\(I_{\rm int}\) denotes a local, parametrically weak coupling to the original
corner sector; its role in fixing the endpoint-to-defect-region map is
described below.  For brevity, the notation \(Z^{\rm extra}_{\Sigma}\) used
below includes this weak interaction.  This extra sector is not the standard corner CFT that appears as the boundary
dual of undeformed wedge holography.  That corner theory is already encoded
by the wedge RT area \(A_W/(4G_W)\).  Equation~\eqref{eq:defectaction}
instead defines a defect-deformed wedge model in which the original corner
theory is coupled to an extra localized quantum sector.  Only the matter
entropy associated with this extra sector, including perturbative interaction
corrections, is added to the generalized entropy, so no double counting is
involved.

\emph{Why the defect entropy is not computed by the wedge RT area.---}
The additional defect CFT introduced in Eq.~\eqref{eq:defectaction} should not
be interpreted as a deformation of the same large-\(N\) corner theory whose
entropy is already computed by the wedge RT surface.  If that were the
interpretation, the deformation would define a new single wedge-holographic
theory, the classical bulk would be replaced by a deformed wedge
\(W_{\rm def}\), and the entropy would have to be computed by an RT/QES
surface in \(W_{\rm def}\).

The construction considered here is instead semiclassical.  The wedge bulk
describes the original gravitational/corner sector, while the added defect CFT
is a distinct additional localized matter sector, weakly coupled to the
original corner theory and to the induced metric \(\sigma_{ij}\) on the
codimension-two corner.  Here ``distinct'' means that its Hilbert-space degrees
of freedom and entropy are not already encoded by the wedge area; it does not
mean that the two corner sectors are dynamically decoupled.  The total path
integral is therefore of the form
\begin{equation}
        Z_{\rm total}
        =
        \int \mathcal{D}g\,
        e^{-I_{\rm wedge}[g]}\,
        Z^{\rm extra}_{\Sigma}[\sigma(g)] ,
\label{eq:total-path-integral-defect}
\end{equation}
or equivalently
\begin{equation}
        I_{\rm eff}[g]
        =
        I_{\rm wedge}[g]
        -
        \log Z^{\rm extra}_{\Sigma}[\sigma(g)] .
\label{eq:effective-action-defect}
\end{equation}
The added sector backreacts on the wedge geometry through its stress tensor,
\begin{equation}
        \frac{\delta I_{\rm wedge}}{\delta g_{\mu\nu}}
        +
        \frac{1}{2}
        \sqrt{\sigma}\,
        \langle T^{ij}_{\Sigma}\rangle_{\rm extra}
        \frac{\delta \sigma_{ij}}{\delta g_{\mu\nu}}
        =
        0 ,
\label{eq:defect-backreaction-equation}
\end{equation}
but its entanglement entropy remains a matter-entropy contribution to the
generalized entropy.  Thus the relevant QES functional is
\begin{equation}
        S_{\rm gen}
        =
        \frac{A_W[g_{\rm cl}](\Gamma_W)}{4G_W}
        +
        S^{\rm extra}_{\Sigma}(A_{\Sigma})
        +\cdots .
\label{eq:qes-functional-extra-defect}
\end{equation}
The first term is the gravitational area contribution of the wedge sector.
The second term is the von Neumann entropy of the additional localized defect
degrees of freedom.  It is not computed by the wedge RT surface, because it is
not the entropy of the original corner CFT already dual to the wedge.

If the added defect CFT is itself holographic, then its partition function may
be represented by an auxiliary bulk,
\begin{equation}
        Z^{\rm extra}_{\Sigma}[\sigma]
        \simeq
        \int_{\partial W_{\rm aux}=\Sigma}
        \mathcal{D}g_{\rm aux}\,
        e^{-I_{\rm aux}[g_{\rm aux};\sigma]} .
\label{eq:aux-partition-defect}
\end{equation}
Consequently its entropy can be computed by the auxiliary RT/QES prescription,
\begin{equation}
        S^{\rm extra}_{\Sigma}(A_{\Sigma})
        =
        \underset{\gamma_{\Sigma}\sim A_{\Sigma}}{\rm ext}
        \left[
        \frac{A_{\rm aux}(\gamma_{\Sigma})}{4G_{\rm aux}}
        +
        S^{\rm bulk}_{\rm aux}
        \right].
\label{eq:aux-qes-defect}
\end{equation}
The auxiliary bulk is therefore not a replacement for the wedge bulk.  It is
an ``integrated-in'' holographic representation of the matter entropy
\(S^{\rm extra}_{\Sigma}\).  The wedge bulk computes the gravitational area
term, while the auxiliary bulk computes the entropy of the added matter sector.

This distinction is the same as in the ordinary QES formula: matter fields can
gravitate and backreact on the classical geometry, but their entropy is not
thereby absorbed into the area term.  Backreaction is controlled by the
one-point function \(\langle T^{ij}_{\Sigma}\rangle_{\rm extra}\), whereas the
entropy contribution is controlled by the reduced density matrix of the defect
sector.  These are logically distinct.  In particular, a CFT vacuum can have
nonzero interval entanglement entropy,
\begin{equation}
        S^{\rm extra}_{\Sigma}
        =
        \frac{c_{\Sigma}}{3}
        \log\frac{\ell_{\Sigma}}{\epsilon_{\Sigma}},
\label{eq:defect-vacuum-entropy}
\end{equation}
while having vanishing stress tensor,
\begin{equation}
        \langle T^{ij}_{\Sigma}\rangle_{\rm extra}=0 .
\label{eq:defect-vacuum-stress}
\end{equation}
Thus an \(O(c_{\Sigma})\) defect entropy contribution can affect the QES
condition even in a regime where the leading classical wedge geometry is
unchanged.

In the probe-defect regime one has, schematically,
\begin{equation}
        G_W \langle T^{ij}_{\Sigma}\rangle_{\rm extra}
        \ll
        \mathcal{R}^{ij}[g_{\rm cl}] ,
\label{eq:probe-defect-schematic}
\end{equation}
so that
\begin{equation}
        g_{\rm cl}
        =
        g^{(0)}
        +
        O\!\left(G_W\langle T_{\Sigma}\rangle_{\rm extra}\right).
\label{eq:probe-defect-metric}
\end{equation}
To leading order the wedge area is evaluated on the undeformed Neumann wedge,
whereas the defect entropy remains in the generalized entropy functional:
\begin{equation}
        S_{\rm gen}
        =
        \frac{A_W[g^{(0)}](\Gamma_W)}{4G_W}
        +
        S^{\rm extra}_{\Sigma}(A_{\Sigma})
        +
        O\!\left(G_W\langle T_{\Sigma}\rangle_{\rm extra}\right).
\label{eq:probe-defect-qes}
\end{equation}
The endpoint condition is therefore
\begin{equation}
        \frac{1}{4G_W}\,
        \partial_{z_a}A_W[g^{(0)}]
        +
        \partial_{z_a}S^{\rm extra}_{\Sigma}
        =
        0
\label{eq:probe-defect-endpoint}
\end{equation}
at leading order.  A fully backreacted treatment would require solving the
coupled wedge-plus-defect equations and then using the same generalized
entropy formula with \(g_{\rm cl}\) replacing \(g^{(0)}\).

For notational brevity, in the following we write
\(S_\Sigma\equiv S^{\rm extra}_\Sigma\) whenever no confusion can arise.

For an island or radiation partition whose endpoint on the branes is
\(z_a\), let the added defect CFT be entangled over a region
\(A_\Sigma(z_a)\) on \(\Sigma\). This region need not be the whole corner: the QES problem is defined for a
specified radiation/island partition, and the endpoint position selects the
defect subregion associated with that partition.  The whole-corner entropy of
the original wedge dual remains represented by the wedge RT area;
\(A_\Sigma(z_a)\) denotes only the subregion of the additional localized
defect matter sector whose entropy is included in the generalized entropy.
The generalized entropy is
\begin{equation}
 S_{\rm gen}(z_a)
 =
 {A_W(\Gamma_W(z_a))\over 4G_W}
 +
 S_\Sigma(A_\Sigma(z_a)) .
\label{eq:Sgenmatter}
\end{equation}

The relation between the brane endpoint and the defect region is fixed by a
microscopic scale-matching coupling at the corner, rather than by choosing an
entropy function after the fact.  In the corner description, a bottom-up
microscopic realization is obtained from the local interaction
\begin{equation}
\begin{split}
 I_{\rm int}
 &={}
 \lambda_\ast
 \int_{\Sigma}d^{d-1}x\,\sqrt{-\sigma}\,
 \mathcal O_{0}(x)\mathcal O_{\rm extra}(x),\\
 \Delta_{0}+\Delta_{\rm extra}
 &={}
 d-1 .
\end{split}
\label{eq:corner-scale-coupling}
\end{equation}
where \(\mathcal O_{0}\) belongs to the original corner sector and
\(\mathcal O_{\rm extra}\) belongs to the added defect CFT.  The dimension
condition makes the double-trace interaction classically marginal.  We assume
that \(\lambda_\ast\) is tuned to a conformal fixed point, or remains
sufficiently close to one over the range of scales probed by the QES; exact
marginality is not inferred from engineering dimensions alone.  Since the
interaction is local on the common induced geometry \((\Sigma,\sigma)\), the
two sectors are coarse grained at the same proper Wilsonian scale \(\mu\).
Their holographic radial dictionaries can be written as
\begin{equation}
 \mu={\alpha_W\over z},
 \qquad
 \mu={\alpha_{\rm aux}\over\zeta},
\label{eq:common-wilsonian-scale}
\end{equation}
where \(\alpha_W\) and \(\alpha_{\rm aux}\) encode the normalization of the
two radial coordinates.  Matching the scale on the two sides therefore gives
\begin{equation}
 \zeta=\frac{\alpha_{\rm aux}}{\alpha_W}\,z .
\label{eq:radial-scale-matching}
\end{equation}

The same identification follows directly in the replica construction.  A
candidate wedge endpoint at \(z=z_a\) fixes the Wilsonian scale
\(\mu_a=\alpha_W/z_a\) at which the replica cut of the combined corner system
is resolved.  Locality of \eqref{eq:corner-scale-coupling} transfers that cut
to the added defect sector at the same \(\mu_a\); its support defines the
boundary of the corresponding region \(A_\Sigma(z_a)\),
\begin{equation}
 \mathfrak a:\quad
 (r_a,z_a,\vec y_0)\in\partial\Gamma_W\cap Q_a
 \longmapsto
 \partial A_\Sigma(z_a)\subset\Sigma .
\label{eq:anchoring-map}
\end{equation}
At a conformal fixed point the size of this region obeys
\begin{equation}
 \mu_a\,\ell_\Sigma(z_a)=\kappa ,
\label{eq:replica-scale-matching}
\end{equation}
with \(\kappa>0\) fixed by the microscopic interaction and the regulator
convention.  Thus \(\ell_\Sigma\propto z_a\) in the vacuum.  If the added
defect sector is holographic, the same common-scale condition fixes the
turning point of its auxiliary RT surface to be
\begin{equation}
 \zeta_\ast(z_a)=\chi z_a,
 \qquad
 \chi\equiv{\alpha_{\rm aux}\over\alpha_W}>0 .
\label{eq:turning-map}
\end{equation}
The map is therefore the radial expression of a local Wilsonian matching
condition, and \(\chi\) is determined by the relative normalization of the
two holographic RG schemes rather than chosen to obtain the desired sign.
This gives a concrete bottom-up microscopic realization of the map; it is not
a claim that every possible defect deformation must generate the same
normalization \(\chi\).  Pushing the wedge endpoint toward the horizon lowers
\(\mu_a\), enlarges the replica region in the defect theory, and moves the
auxiliary RT surface deeper into \(W_{\rm aux}\).

We stress that the weak interaction is not invoked to perturbatively generate
the magnitude of \(S_\Sigma^{\rm extra}\).  Its role is to specify which defect
subalgebra, and hence which replica cut, is paired with a candidate wedge
endpoint.  Once \(A_\Sigma(z_a)\) is fixed, the intrinsic entropy of a
large-\(c_\Sigma\) defect sector can be \(O(c_\Sigma)\), whereas the
interaction-induced one-point function, geometric backreaction, and
self-energy corrections remain parametrically small.  There is therefore no
conflict between a weak scale-matching interaction and a leading defect
entropy in the generalized entropy.

The interaction in \eqref{eq:corner-scale-coupling} is a scalar under the
common corner diffeomorphisms and preserves the conserved diagonal
stress-tensor Ward identity.  If both sectors are represented holographically,
the coupling can mix their spin-two channels and can give a mass to the
orthogonal combination, but conservation of the diagonal stress tensor leaves
one massless spin-two combination.  In the weak-coupling limit this mode is
continuously connected to the positive-norm Neumann zero mode
\cite{Aharony:2006hz}.  We take \(\lambda_\ast\) sufficiently weak that the
stress-tensor backreaction and the remaining self-energy corrections to the
Neumann KK operator are perturbative.

For example, if the added defect theory is a two-dimensional holographic
CFT with auxiliary BTZ/AdS\(_3\) geometry
\begin{align}
 ds_{\rm aux}^2
 &=
 {1\over \zeta^2}
 \left(
 {d\zeta^2\over f_{\rm aux}(\zeta)}
 -f_{\rm aux}(\zeta)dt^2+dy^2
 \right),
\nonumber\\
 f_{\rm aux}(\zeta)
 &=1-{\zeta^2\over \zeta_h^2},
\label{eq:aux-btz}
\end{align}
then an equal-time interval \(A_\Sigma(z_a)\) of length
\(\ell_\Sigma(z_a)\) has an RT geodesic with turning point
\(\zeta_\ast(z_a)\).  The standard relation gives
\begin{equation}
 \ell_\Sigma(z_a)
 =
 2\zeta_h\,\operatorname{arctanh}
 \left({\chi z_a\over \zeta_h}\right),
 \qquad
 0<\chi z_a<\zeta_h .
\label{eq:ell-map-btz}
\end{equation}
Therefore
\begin{equation}
 {d\ell_\Sigma\over dz_a}
 =
 {2\chi\over 1-(\chi z_a/\zeta_h)^2}
 >0 .
\label{eq:ell-positive-derived}
\end{equation}
In the auxiliary vacuum limit \(\zeta_h\to\infty\), this reduces to
\begin{equation}
 \ell_\Sigma(z_a)=2\chi z_a .
\label{eq:ell-map-vacuum}
\end{equation}
Thus the monotonicity condition used below is not an independent sign
assumption; it follows from the concrete radial anchoring map
\eqref{eq:turning-map}.

More generally, if the defect CFT is holographic, then
\begin{equation}
 S_\Sigma(A_\Sigma)
 =
 {A_{\rm aux}(\gamma_\Sigma)\over 4G_{\rm aux}}
 +
 S_{\rm aux}^{\rm bulk}
 +\cdots ,
\label{eq:auxRT}
\end{equation}
where \(\gamma_\Sigma\subset W_{\rm aux}\) is the RT surface in the
auxiliary \(d\)-dimensional bulk dual to the added \((d-1)\)-dimensional
defect theory.  The total generalized entropy is then
\begin{equation}
 S_{\rm gen}(z_a)
 =
 {A_W(\Gamma_W(z_a))\over 4G_W}
 +
 {A_{\rm aux}(\gamma_\Sigma(z_a))\over 4G_{\rm aux}}
 +\cdots .
\label{eq:Sgenaux}
\end{equation}

The computational framework is therefore as follows.  For each trial endpoint
\(z_a\), one first computes the endpoint-reduced wedge area
\(A_W[g^{(0)}](z_a)\) in the healthy Neumann wedge background.  The common-scale replica map
then assigns to the same endpoint a nested defect region \(A_\Sigma(z_a)\),
whose entropy is computed directly in the defect CFT, or by the auxiliary
RT/QES prescription when the defect CFT is holographic.  The endpoint is then
fixed by extremizing
\begin{equation}
\begin{split}
        S_{\rm gen}(z_a)
        ={}&
        {A_W[g^{(0)}](z_a)\over 4G_W}
        +
        S_\Sigma^{\rm extra}[A_\Sigma(z_a);\sigma^{(0)}]\\&
        +
        O\!\left(G_W\langle T_\Sigma\rangle_{\rm extra}\right).
\label{eq:computational-framework}
\end{split}
\end{equation}
This expression is the leading term in the probe-defect expansion.  In a
fully backreacted solution one should replace \(g^{(0)}\) and
\(\sigma^{(0)}\) by the classical solution \(g_{\rm cl}\) and induced metric
\(\sigma_{\rm cl}\) sourced by \(\langle T^{ij}_\Sigma\rangle_{\rm extra}\):
\begin{equation}
        S_{\rm gen}
        =
        {A_W[g_{\rm cl}](\Gamma_W)\over 4G_W}
        +
        S_\Sigma^{\rm extra}[A_\Sigma;\sigma_{\rm cl}]
        +\cdots .
\label{eq:backreacted-qes-computational}
\end{equation}
Consistency with the deformed wedge therefore means that the backreaction
corrections are perturbatively small in the regime considered; it does not
mean that the defect matter entropy is to be replaced by the wedge area term.
Backreaction changes the geometry entering \(A_W\), while the defect entropy
remains the matter-entropy term of the QES functional.

This also clarifies the relation to causal wedge inclusion (CWI).  The CWI argument in the minimal wedge theory applies to the RT surface
computing the entropy of the original corner CFT in a single wedge-holographic
description.  In that sector the conclusion is unchanged: the purely geometric
RT problem selects the horizon-collapsed surface.  The non-horizon surface
discussed here is not a competing classical RT surface for the original corner
CFT.  It is a QES of the enlarged semiclassical system, whose generalized
entropy contains the additional matter term \(S^{\rm extra}_{\Sigma}\).  When
an auxiliary RT surface is used, it computes only the entropy of this added
defect matter sector and must satisfy the usual causal-wedge/entanglement-wedge
consistency conditions of its own auxiliary holographic dual.  Thus the
construction does not violate CWI in the original wedge theory; it changes the
entropy functional by coupling the healthy Neumann wedge sector to additional
quantum matter.

For the full coupled system, the relevant inclusion statement concerns the
quantum entanglement wedge determined by the complete generalized entropy,
not the classical entanglement wedge determined by \(A_W\) alone.  At the
order retained here this statement is implemented by generalized-entropy
maximin, the homology constraint, the nested common-scale map
\(A_\Sigma(z_a)\), and the standard semiclassical energy conditions
\cite{Wall:2012uf,Almheiri:2019psf}.  A fully backreacted nested-holographic
completion would geometrize both contributions in
Eq.~\eqref{eq:backreacted-qes-computational}; it would not identify the added
matter entropy with the original wedge area.  Consequently the minimal-wedge
CWI argument cannot force the enlarged-system QES to the horizon once
\(\partial_{z_a}S_\Sigma^{\rm extra}\neq0\).

\begin{figure}[t]
\centering
\resizebox{0.96\columnwidth}{!}{%
\begin{tikzpicture}[scale=1.0,>=stealth]
  
  \draw[blue!70!black,line width=1.2pt] (-2.2,0) -- (-1.5,3.0);
  \draw[blue!70!black,line width=1.2pt] (2.2,0) -- (1.5,3.0);
  \node[blue!70!black] at (-2.35,2.55) {$Q_1$};
  \node[blue!70!black] at (2.35,2.55) {$Q_2$};
  \node at (0,2.45) {$W$};
  \node at (0,-0.25) {$\Sigma$};
  \fill (0,0) circle (1.5pt);
  \draw[dashed,black] (-1.25,0.9) .. controls (-0.5,0.55) and (0.5,0.55) .. (1.25,0.9);
  \node at (0,0.72) {\scriptsize horizon};
  \draw[red,line width=1.2pt] (-1.82,1.65) .. controls (-0.7,1.35) and (0.7,1.35) .. (1.82,1.65);
  \fill[red] (-1.82,1.65) circle (2.5pt);
  \fill[red] (1.82,1.65) circle (2.5pt);
  \node[red] at (-1.95,1.9) {$z_1$};
  \node[red] at (1.95,1.9) {$z_2$};
  \node[red] at (0,1.55) {$\Gamma_W$};
  \draw[->,red,dashed] (1.95,1.65) .. controls (3.0,1.95) and (3.4,2.55) .. (4.0,2.55);
  \node at (3.05,2.15) {\scriptsize $z_a\mapsto A_\Sigma(z_a)$};

  \draw[green!50!black,line width=1.1pt] (5.3,1.6) ellipse (1.2 and 1.35);
  \draw[green!50!black,line width=1.2pt] (4.55,2.55) -- (6.05,2.55);
  \fill[green!50!black] (4.55,2.55) circle (2.2pt);
  \fill[green!50!black] (6.05,2.55) circle (2.2pt);
  \draw[green!50!black,dashed,line width=1.1pt] (4.55,2.55) .. controls (4.85,1.55) and (5.75,1.55) .. (6.05,2.55);
  \node[green!50!black] at (5.3,3.18) {$W_{\rm aux}$};
  \node[green!50!black] at (5.3,2.78) {$A_\Sigma(z_a)$};
  \node[green!50!black] at (5.3,1.45) {$\gamma_\Sigma$};
  
  \node[draw=red,rounded corners,align=center,text=red] at (-0.15,-0.95) {\scriptsize wedge area: $\partial_{z_a}A_W<0$\\[-1mm]\scriptsize toward horizon};
  \node[draw=green!50!black,rounded corners,align=center,text=green!50!black] at (5.25,-0.95) {\scriptsize auxiliary area: $\partial_{z_a}A_{\rm aux}>0$\\[-1mm]\scriptsize if $A_\Sigma$ grows};
\end{tikzpicture}}
\caption{Defect-assisted wedge island.  The wedge RT surface $\Gamma_W$ has endpoints $z_a$ on the branes.  In the undeformed wedge the area decreases as the endpoints move toward the horizon, so the horizon is selected.  The added holographic defect CFT assigns to the same endpoint a defect region $A_\Sigma(z_a)$ whose RT surface $\gamma_\Sigma$ lies in an auxiliary bulk $W_{\rm aux}$.  If $A_\Sigma$ grows in the horizon direction, the auxiliary area gives the opposite variation and the joint extremality condition can stabilize a non-horizon island saddle.}
\label{fig:mechanism}
\end{figure}
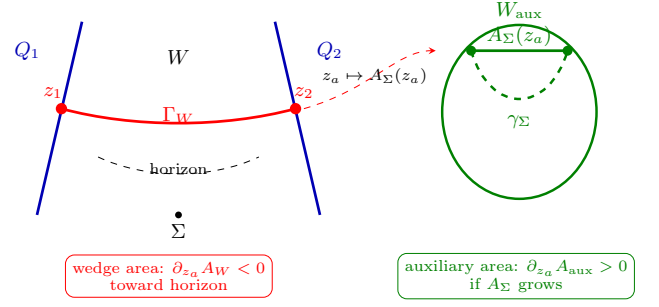

Varying the common endpoint gives
\begin{equation}
 {1\over 4G_W}{\partial A_W\over \partial z_a}
 +
 {\partial S_\Sigma\over \partial z_a}
 =0,
\label{eq:boundarymatter}
\end{equation}
or, in the holographic defect limit,
\begin{equation}
 {1\over 4G_W}{\partial A_W\over \partial z_a}
 +
 {1\over 4G_{\rm aux}}{\partial A_{\rm aux}\over \partial z_a}
 =0 .
\label{eq:boundaryaux}
\end{equation}
Equation~\eqref{eq:boundaryaux} is the central point.  The pure wedge
boundary condition \(\partial_{z_a}A_W=0\), equivalently orthogonality at the
brane, is replaced by a quantum extremality condition involving a second
healthy entropy functional.  Therefore the surface need not end
orthogonally on the brane; its endpoint is instead fixed by the balance
between the wedge area and the defect entropy, as illustrated in
Fig.~\ref{fig:mechanism}.

\emph{Maximin interpretation.---} Equation~\eqref{eq:boundaryaux} should be understood as the local endpoint form of a maximin problem for the full generalized entropy.  This point is important because the minimal wedge obstruction is not only that the classical RT surface is driven to the horizon, but also that the purely geometric problem fails to select a nontrivial radiation partition.  This is closely analogous to the puzzle in three-dimensional wedge holography.  If the branes are treated as rigid, the classical RT functional admits a degenerate family of equal-area surfaces, so the entanglement wedge is not uniquely defined.  Including the radion mode gives the effective Jackiw--Teitelboim (JT) description \cite{Jackiw:1984je,Teitelboim:1983ux}, and the maximin prescription selects a definite HRT surface \cite{Hubeny:2007xt,Geng:2022slq,Wall:2012uf}.

In the present problem the added defect entropy plays the corresponding variational role.  On each Cauchy slice one should minimize
\begin{equation}
 {\cal S}
 =
 \frac{A_W(\Gamma_W)}{4G_W}
 +
 \frac{A_{\rm aux}(\gamma_\Sigma)}{4G_{\rm aux}}
 +\cdots ,
\label{eq:maximin-functional}
\end{equation}
subject to the homology constraint and to the anchoring relation $z_a\mapsto A_\Sigma(z_a)$, and then maximize over Cauchy slices.  The pure wedge problem corresponds to dropping the second term in Eq.~\eqref{eq:maximin-functional}; the resulting endpoint condition is the orthogonality condition $z'_a=0$, which selects the horizon.  With the defect entropy included, the maximin functional instead selects an extremum of the combined entropy.  Thus the non-horizon island surface is not obtained by imposing a non-geometric boundary condition by hand.  It is selected by the standard generalized-entropy prescription applied to the enlarged wedge-plus-defect system.

\emph{Opposite sign without ghosts.---} The sign needed for a non-horizon saddle is not a negative central charge.  It is a monotonicity condition.  In the convention where increasing $z_a$ moves the wedge endpoint toward the horizon, the wedge area satisfies
\begin{equation}
 \frac{\partial A_W}{\partial z_a}<0 .
\label{eq:wedgevariation}
\end{equation}
A competing defect sector must satisfy
\begin{equation}
 \frac{\partial S_\Sigma}{\partial z_a}>0,
\qquad\text{or}\qquad
 \frac{\partial A_{\rm aux}}{\partial z_a}>0 .
\label{eq:oppositesign}
\end{equation}
This can occur with $G_{\rm aux}>0$ and a positive defect central charge.  Suppose the endpoint determines a nested family of defect regions with size $\ell_\Sigma(z_a)$ and $d\ell_\Sigma/dz_a>0$.  For a two-dimensional holographic defect CFT,
\begin{equation}
 S_\Sigma(z_a)=\frac{c_\Sigma}{3}\log\frac{\ell_\Sigma(z_a)}{\epsilon},
\label{eq:2dcft}
\end{equation}
so that
\begin{equation}
 \frac{\partial S_\Sigma}{\partial z_a}
 =\frac{c_\Sigma}{3}\frac{\ell_\Sigma'(z_a)}{\ell_\Sigma(z_a)}>0
\label{eq:positivevariation}
\end{equation}
whenever $c_\Sigma>0$ and $\ell_\Sigma'(z_a)>0$.  In the auxiliary AdS$_3$ description this is exactly the statement that a geodesic length increases when the boundary interval grows.  In higher-dimensional holographic defect theories the same conclusion follows for nested regions from entanglement wedge nesting, up to the standard local counterterms.

A sufficient existence criterion follows immediately.  Define
\begin{equation}
 F(z_a)=\frac{1}{4G_W}\partial_{z_a}A_W
 +\frac{1}{4G_{\rm aux}}\partial_{z_a}A_{\rm aux} .
\label{eq:F}
\end{equation}
If $F(z_0)<0$ for some endpoint away from the horizon and $F(z_1)>0$ sufficiently close to the horizon, then by continuity there is a stationary point $z_\ast\in(z_0,z_1)$.  When $\partial_{z_a}^2S_{\rm gen}(z_\ast)>0$ and the corresponding entropy is below the late-time Hartman-Maldacena entropy, this stationary point is the island saddle.  The Hartman-Maldacena surface grows linearly at late times in the eternal black-string geometry \cite{Hartman:2013qma}, whereas the island entropy is time independent; hence a Page transition follows once the non-horizon island saddle exists and has finite generalized entropy.

\emph{Local endpoint model.---}
To make the mechanism explicit while staying close to the black-string
variational problem, we expand the endpoint-reduced wedge area near the
horizon-collapsed saddle.  Let
\begin{equation}
 u=1-z_a,\qquad u=0
\label{eq:udef}
\end{equation}
denote the horizon endpoint.  For each fixed endpoint \(z_a\), define
\begin{equation}
 {\cal A}_W(z_a)=
 \min_{\Gamma_W:\,\partial\Gamma_W=z_a} A_W(\Gamma_W) .
\label{eq:reduced-area}
\end{equation}
The horizon is a one-sided minimum of the undeformed wedge problem.  To see
the leading endpoint dependence directly from Eq.~\eqref{eq:wedgearea}, set
\begin{equation}
 z(r)=1-u\,Y(r),
\label{eq:z-uY}
\end{equation}
where \(Y(r)\geq0\) encodes the shape of the endpoint deformation.  Since
\begin{equation}
 f(z)=1-z^{d-1}=(d-1)uY+O(u^2),
\label{eq:f-local}
\end{equation}
the endpoint-reduced area has the expansion
\begin{equation}
 { {\cal A}_W(u)\over 4G_W}
 =
 S_{\rm hor}
 +
 \mu_W u
 +
 {\kappa_W\over 2}u^2
 +
 O(u^3),
\label{eq:local-wedge-expansion}
\end{equation}
with
\begin{align}
 \mu_W
 &=
 {V\over 4G_W}
 \min_Y
 \int_{-\rho_1}^{\rho_2}dr\,
 \cosh^{d-2}r
 \nonumber\\
&\quad\times
 \left[
 (d-2)Y
 +
 {\cosh^2 r\over 2(d-1)}
 {Y'^2\over Y}
 \right] .
\label{eq:muW}
\end{align}
Since \(Y\geq0\) and the endpoint displacement is nontrivial, the
integrand is non-negative and not identically zero; hence
\begin{equation}
 \mu_W>0 .
\end{equation}
Thus the pure wedge area increases when the endpoint is displaced away
from the horizon.  The coefficient \(\kappa_W\) is the next endpoint
variation of the reduced area.  Its explicit value can be extracted by
continuing the same expansion of Eq.~\eqref{eq:wedgearea}, but it is not
needed for the leading local existence criterion.

The added defect theory is taken to be a unitary two-dimensional CFT whose
entangling interval has length \(\ell_\Sigma(z_a)\).  Its entropy is
\begin{equation}
 S_\Sigma(z_a)
 =
 {c_\Sigma\over 3}
 \log {\ell_\Sigma(z_a)\over \epsilon_\Sigma},
 \qquad c_\Sigma>0 .
\label{eq:local-defect-entropy}
\end{equation}
With the radial corner anchoring map \eqref{eq:turning-map}, the defect
interval grows when the wedge endpoint is pushed toward the horizon.  For
the auxiliary BTZ/AdS\(_3\) example \eqref{eq:ell-map-btz},
\begin{equation}
 {d\ell_\Sigma\over dz_a}
 =
 {2\chi\over 1-(\chi z_a/\zeta_h)^2}>0 .
\label{eq:ell-growth-local}
\end{equation}
Thus the required monotonicity follows from the concrete anchoring map,
rather than from an independent sign assumption.

Near \(u=0\), the defect entropy expands as
\begin{equation}
 S_\Sigma(u)
 =
 S_\Sigma^{(h)}
 -
 \beta u
 +
 {\lambda_\Sigma\over 2}u^2
 +
 O(u^3),
\label{eq:defect-local-expansion}
\end{equation}
where
\begin{equation}
 \beta
 =
 {c_\Sigma\over 3}
 \left.
 {d\log\ell_\Sigma\over dz_a}
 \right|_{z_a=1}
 >0 .
\label{eq:beta-positive-general}
\end{equation}
For the explicit BTZ/AdS\(_3\) anchoring map this gives
\begin{equation}
 \beta
 =
 {c_\Sigma\over 3}\,
 {\chi\over
 \zeta_h\!\left[1-(\chi/\zeta_h)^2\right]
 \operatorname{arctanh}(\chi/\zeta_h)}
 >0 .
\label{eq:beta-positive}
\end{equation}
In the auxiliary vacuum limit \(\zeta_h\to\infty\), this reduces to
\begin{equation}
 \beta={c_\Sigma\over 3}.
\label{eq:beta-vacuum}
\end{equation}

The combined generalized entropy is therefore
\begin{equation}
 S_{\rm gen}(u)
 =
 S_{\rm hor}+S_\Sigma^{(h)}
 +
 (\mu_W-\beta)u
 +
 {1\over 2}\kappa_{\rm eff}u^2
 +
 O(u^3),
\label{eq:local-Sgen}
\end{equation}
with
\begin{equation}
 \kappa_{\rm eff}\equiv \kappa_W+\lambda_\Sigma .
\label{eq:kappaeff}
\end{equation}
The coefficient \(\lambda_\Sigma\) is not assumed to be positive.  For a
logarithmic defect entropy, its sign depends on the second derivative of
the anchoring function \(\ell_\Sigma(z_a)\).  The required condition is the
stability condition
\begin{equation}
 \kappa_{\rm eff}>0 .
\label{eq:kappaeff-positive}
\end{equation}

The defect-assisted saddle appears when
\begin{equation}
 \beta>\mu_W .
\label{eq:beta-threshold}
\end{equation}
Equivalently, for the explicit auxiliary BTZ/AdS\(_3\) anchoring map,
\begin{equation}
 {c_\Sigma\over 3}\,
 {\chi\over
 \zeta_h\!\left[1-(\chi/\zeta_h)^2\right]
 \operatorname{arctanh}(\chi/\zeta_h)}
 >
 \mu_W .
\label{eq:explicit-threshold}
\end{equation}
In the auxiliary vacuum limit this reduces to
\begin{equation}
 {c_\Sigma\over 3}>\mu_W .
\label{eq:vacuum-threshold}
\end{equation}
When Eq.~\eqref{eq:beta-threshold} holds, the horizon is no longer a local
minimum of the full generalized entropy.  The stationary point is
\begin{equation}
 u_\ast
 =
 {\beta-\mu_W\over \kappa_{\rm eff}}
 +
 O\!\left((\beta-\mu_W)^2\right),
 \qquad
 z_\ast=1-u_\ast<1 .
\label{eq:local-ustar}
\end{equation}
It is a stable non-horizon QES, provided the displacement remains within
the regime of the endpoint expansion.  Thus the controlled regime is
\begin{equation}
 0<u_\ast
 =
 {\beta-\mu_W\over \kappa_{\rm eff}}
 \ll 1 ,
\label{eq:local-validity}
\end{equation}
which is the near-threshold weak endpoint-displacement regime.  In this
regime the QES is displaced away from the horizon but remains close enough
to the horizon-collapsed saddle for the local expansion to be reliable.

Moreover,
\begin{equation}
 S_{\rm gen}(u_\ast)-S_{\rm gen}(0)
 =
 -
 {(\beta-\mu_W)^2\over 2\kappa_{\rm eff}}
 +
 O\!\left((\beta-\mu_W)^3\right)
 <0 .
\label{eq:local-dominance}
\end{equation}
Thus the defect-assisted saddle is strictly preferred over the
horizon-collapsed saddle within the controlled local regime.  If
\(\beta-\mu_W\) is too large, the saddle can move outside the range of the
endpoint expansion, and the full joint RT problem must be solved.

The comparison with the no-island Hartman--Maldacena saddle is standard.
The local endpoint analysis establishes a finite, time-independent
defect-assisted island entropy \(S_{\rm island}\).  By contrast, the
Hartman--Maldacena surface grows at late times,
\begin{equation}
 S_{\rm HM}(t)
 =
 S_{\rm HM}(0)+v_{\rm HM}t+\cdots,
 \qquad v_{\rm HM}>0 .
\label{eq:hm-growth}
\end{equation}
Therefore, provided the no-island saddle dominates initially,
\(S_{\rm HM}(0)<S_{\rm island}\), the two saddles exchange dominance at
\begin{equation}
 t_{\rm Page}
 =
 {S_{\rm island}-S_{\rm HM}(0)\over v_{\rm HM}} .
\label{eq:toy-page-time}
\end{equation}
For \(t>t_{\rm Page}\), the defect-assisted island saddle dominates.  Thus
the local endpoint model supplies the finite non-horizon island saddle,
while the usual late-time growth of the Hartman--Maldacena surface supplies
the Page transition.

\emph{Health of the construction and KK spectrum.---}
The mechanism above keeps the gravitational wedge sector on the healthy
massless Neumann branch.  The point is stronger than the positivity of
\(c_\Sigma\) and \(G_{\rm aux}\).  Those are necessary consistency conditions
for the added defect sector, but ghost-freedom of the gravitational sector is
instead controlled by the sign of the kinetic term and by the norm of the
Kaluza--Klein modes.  Since no DGP term is introduced, the defect construction
does not add a brane-localized Einstein--Hilbert term with negative
coefficient.  Therefore, in the probe-defect regime, the transverse-traceless
graviton fluctuations obey the same Sturm--Liouville problem as in the
undeformed Neumann wedge.

Let
\begin{equation}
        h_{ij}(x,r)
        =
        \sum_n \psi_n(r)\,h^{(n)}_{ij}(x),
        \quad
        \nabla^i h^{(n)}_{ij}=0,
        \quad
        h^{(n)i}{}_{i}=0 ,
\label{eq:kk-decomposition}
\end{equation}
where \(i,j\) denote directions along the brane slices.  For the
black-string wedge \eqref{eq:blackstring}, the transverse-traceless radial
equation following from the quadratic Einstein action is explicitly
\cite{Hu:2022ymx}
\begin{align}
 -{d\over dr}\!\left(\cosh^{d}r\,{d\psi_n\over dr}\right)
 &=m_n^2\cosh^{d-2}r\,\psi_n,
\label{eq:neumann-kk-problem}\\
 \psi_n'(-\rho_1)&=\psi_n'(\rho_2)=0 .
\label{eq:neumann-kk-bc}
\end{align}
The Sturm--Liouville weight is therefore
\(w(r)=\cosh^{d-2}r>0\).  Multiplying
Eq.~\eqref{eq:neumann-kk-problem} by \(\psi_n\), integrating over the wedge,
and using the Neumann conditions gives the exact identity
\begin{align}
 m_n^2\!\int_{-\rho_1}^{\rho_2}\!dr\,
 \cosh^{d-2}r\,\psi_n^2
 &={}
 \int_{-\rho_1}^{\rho_2}\!dr\,
 \cosh^{d}r\,(\psi_n')^2
 \geq0 .
\label{eq:kk-positive-mass-identity}
\end{align}
The kinetic norm is
\begin{equation}
        \|\psi_n\|^2_{\rm wedge}
        =
        {1\over 16\pi G_W}
        \int_{-\rho_1}^{\rho_2}dr\,
        \cosh^{d-2}r\,\psi_n(r)^2>0
\label{eq:wedge-kk-norm}
\end{equation}
for every normalizable mode.  The zero-eigenvalue solution is
\(\psi_0={\rm constant}\), and hence
\begin{equation}
        m_0^2=0,
        \qquad
        \|\psi_0\|_{\rm wedge}^2>0 .
\label{eq:zero-mode-positive}
\end{equation}
Thus the undeformed Neumann problem has neither a tachyonic spin-two mode nor a
negative-norm KK mode, and its long-range graviton is a positive-norm massless
mode.

This is precisely where the present construction differs from negative DGP.
A brane Einstein term modifies the KK inner product schematically to
\begin{align}
 \|\psi_n\|^2_{\rm DGP}
 ={}&
 {1\over 16\pi G_W}\int dr\,w(r)\psi_n^2
 \nonumber\\
 &+
 \sum_a{\psi_n(r_a)^2\over16\pi G_{{\rm DGP},a}} .
\label{eq:dgp-kk-norm}
\end{align}
On the negative-DGP branch the localized contribution has the wrong sign.
Consequently a spin-two mode can acquire a negative residue even when the
effective low-energy Newton constant is positive.  This is the origin of the
DGP ghost \cite{Miao:2023mui}.  No analogous negative contribution appears
in \eqref{eq:wedge-kk-norm}, because the present construction contains no DGP
kinetic term.

The additional defect CFT couples to the induced corner metric through
\begin{equation}
        \delta I_\Sigma^{\rm extra}
        =
        {1\over 2}
        \int_\Sigma d^{d-1}x\,\sqrt{\sigma}\,
        T_\Sigma^{ij}\,\delta\sigma_{ij}.
\label{eq:defect-stress-coupling}
\end{equation}
At quadratic order, integrating out the defect fields produces a localized
stress-tensor two-point contribution to the effective action,
\begin{equation}
        \Delta I_{\rm eff}^{(2)}
        =
        {1\over 2}
        \int_\Sigma h_{ij}\,
        \Pi_\Sigma^{ij,kl}\,
        h_{kl}.
\label{eq:defect-self-energy}
\end{equation}
For a unitary defect CFT this correlator has a positive spectral
representation,
\begin{align}
 \Pi_\Sigma^{ij,kl}(p)
 ={}&
 \int_0^\infty d\mu^2\,
 \rho_\Sigma(\mu^2)
 {{\cal P}^{ij,kl}(p,\mu)\over p^2+\mu^2-i0},
 \nonumber\\
 &\rho_\Sigma(\mu^2)\geq0 .
\label{eq:positive-spectral-density}
\end{align}
where \({\cal P}^{ij,kl}\) is the spin-two tensor structure.  More generally,
for the coupled original and extra sectors the corresponding spectral-density
matrix is positive semidefinite by unitarity.  The conserved diagonal Ward
identity leaves a zero eigenvalue in the spin-two mass matrix, while an
orthogonal combination may become weakly massive; positive bare kinetic terms
and a positive-semidefinite spectral matrix do not create a negative-residue
state.  In the probe-defect regime the corrections are perturbative,
\begin{equation}
        G_W\,\Pi_\Sigma \ll 1,
\label{eq:probe-self-energy}
\end{equation}
and the surviving massless pole and all KK residues are continuously connected
to those of the healthy Neumann wedge.  Hence no negative-norm pole is
generated in the regime in which the island analysis is performed.

The island saddle is therefore produced by the matter entropy
\(S_\Sigma^{\rm extra}\) in the QES functional, while the gravitational KK
spectrum remains on the ghost-free Neumann branch.  The opposite sign in
\eqref{eq:oppositesign} is a kinematic consequence of how the defect region
depends on the wedge endpoint, not a sign flip in the gravitational action.
Negative DGP gives the desired endpoint force by using an unhealthy
gravitational kinetic term on the brane; the present construction gives the
same type of endpoint force from ordinary positive entanglement entropy.

There are two regimes.  In the probe-defect regime we require
\begin{equation}
 \epsilon_{\rm br}\sim {G_W E_\Sigma\over L^{d-2}}\ll 1,
\label{eq:probe-backreaction}
\end{equation}
so that the background \eqref{eq:blackstring} and the KK problem
\eqref{eq:neumann-kk-problem} are unchanged at leading order, while the defect
entropy remains visible at large central charge.  In a fully backreacted
regime one must solve the coupled wedge-plus-defect background and then analyze
the corresponding coupled fluctuation operator.  The defect stress tensor
should then obey the appropriate averaged null energy condition along the null
generators intersecting the corner,
\begin{equation}
 \int d\lambda\,\langle T^{\Sigma}_{kk}\rangle\geq 0,
\label{eq:defect-anec}
\end{equation}
or, for a holographic defect sector, the classical null energy condition in
\(W_{\rm aux}\).  These assumptions preserve entanglement-wedge nesting and
prevent the backreaction from reversing the monotonicity condition
\(\partial_{z_a}A_{\rm aux}>0\).  The brane tensions are kept on the Neumann
massless branch and the interaction preserves the diagonal Ward identity, so
one positive-norm massless eigenmode survives and is continuously connected to
the Neumann zero mode.  In neither regime is a negative Newton constant,
negative central charge, or negative spin-two kinetic term required.

\emph{Discussion.---} The result clarifies the origin of the massless-island obstruction in minimal wedge holography.  The obstruction is not the existence of a massless graviton by itself.  It is the purely geometric extremization problem, whose free endpoint condition imposes orthogonality and thereby selects the horizon.  Negative DGP changes the endpoint condition but introduces a ghost.  A positive, explicit defect CFT changes the endpoint condition through a matter entropy term.  If the defect entropy is holographic, the same statement is geometrized by an auxiliary RT surface whose area grows in the horizon direction.

The analysis in this Letter is performed in a controlled probe-defect semiclassical regime.  In this regime the weak local coupling fixes the common replica subalgebra and its Wilsonian scale, but does not generate the intrinsic \(O(c_\Sigma)\) defect entropy.  The defect stress tensor and interaction-induced self-energy produce only perturbatively small corrections to the Neumann wedge geometry and KK operator, while the defect entropy enters the generalized entropy at leading order, as in the standard QES prescription.  Therefore a fully backreacted wedge-defect solution is not required for the local existence and perturbative ghost-freedom results established here, although it would be needed to map the complete nonperturbative parameter space of defect-assisted island saddles and to exhibit a single unified geometric dual. It would also be interesting to embed the auxiliary defect sector in a cascaded or nested holographic construction, where all gravitational sectors are healthy and the long-range massless mode remains present.  The central lesson is that massless islands do not require the negative-DGP branch: they require a healthy entropy contribution whose endpoint variation competes with the horizon-minimizing wedge area.

\renewcommand{\bibfont}{\footnotesize}
\setlength{\bibsep}{0pt}
\bibliographystyle{utphys1}
\bibliography{references}

\end{document}